\documentclass{article}
\usepackage{epsf}
\usepackage{emulateapj}
\input{psfig}
\usepackage{mathptm}
\begin{document}
\normalsize

\title{A Black Hole of $>6$ M$_{\odot}$ in the X-ray Nova
XTE~J1118+480$^{1}$}

\author{J. E. McClintock$^{2}$, M. R. Garcia$^{2}$, N. Caldwell$^{3}$,
E. E. Falco$^{3}$, P. M. Garnavich$^{4}$, P. Zhao$^{2}$}

\begin{abstract}

Observations of the quiescent X-ray nova XTE J1118+480 with the new
6.5 m MMT have revealed that the velocity amplitude of the dwarf
secondary is 698~$\pm$~14~km~s$^{-1}$ and the orbital period of the
system is $0.17013~\pm~0.00010$~d.  The implied value of the mass
function, f(M)~=~$6.00~\pm~0.36$~M$_{\odot}$, provides a hard lower
limit on the mass of the compact primary that greatly exceeds the
maximum allowed mass of a neutron star ($\sim$~3~M$_{\odot}$).  Thus
we conclude that the compact primary is a black hole.  Among the
eleven dynamically established black-hole X-ray novae, the large mass
function of XTE J1118+480 is rivaled only by that of V404~Cyg.  We
estimate that the secondary supplies 34\%~$\pm$~8\% of the total light
at 5900~\AA\ and that its spectral type is in the range K5V to M1V.  A
double-humped I-band light curve is probably due to ellipsoidal
modulation, although this interpretation is not entirely secure
because of an unusual 12-minute offset between the spectroscopic and
photometric ephemerides.  Assuming that the light curve is
ellipsoidal, we present a provisional analysis which indicates that
the inclination of the system is high and the mass of the black hole
is correspondingly modest (M$_{1}$~$\lesssim$~10~M$_{\odot}$).  The
broad Balmer emission lines (FWHM~=~2300--2900~km~s$^{-1}$) also
suggest a high inclination.  For the range of spectral types given
above, we estimate a distance of 1.8~$\pm$~0.6~kpc.

\end{abstract}


\setcounter{footnote}{4}

\footnotetext[1]{Observations reported here were obtained at the MMT
Observatory, a facility operated jointly by the University of Arizona
and the Smithsonian Institution.}

\footnotetext[2]{Harvard-Smithsonian Center for Astrophysics, 60
Garden Street, Cambridge, MA 02138; jem@cfa.harvard.edu,
mgarcia@cfa.harvard.edu, pzhao@cfa.harvard.edu.}

\footnotetext[3]{Smithsonian Institution, F. L. Whipple Observatory,
PO Box 97, 670 Mt. Hopkins Road, Amado, AZ 85645;
caldwell@flwo99.sao.arizona.edu, falco@cfa.harvard.edu.}

\footnotetext[4]{Physics Department, University of Notre Dame, Notre
Dame, IN 46556; pgarnavi@nd.edu.}

\normalsize

\section{INTRODUCTION}

The X-ray nova XTE J1118+480 was discovered with the RXTE All-Sky
Monitor on 2000 March 29 (Remillard et al. 2000).  In outburst the
optical counterpart brightened by about 6 mag to V $\approx$~13
(Uemura et al. 2000).  Extensive optical data in outburst reveal that
the orbital period is $\approx$~4.1~hr (Patterson 2000; Uemura et
al. 2000; Garcia et al. 2000; Dubus et al. 2000).  XTE~J1118+480 has
one truly exceptional attribute: Its very high galactic latitude,
b~=~+62$^{\rm o}$, and its correspondingly low reddening, E(B-V)
$\approx$ 0.013 mag (N$_{H} \approx 1.0 \times 10^{20}$~cm$^{-2}$;
Hynes et al. 2000), make it the least reddened of all known X-ray
binaries.

Including XTE J1118+480, very strong evidence now exists for black
hole primaries in eleven X-ray novae (McClintock 1998; Filippenko et
al. 1999; Orosz et al. 2000).  Since XTE~J1118+480 was known to be
optically bright in quiescence (R $\approx$ 18.8; Uemura et al.
2000), the source appeared to be a good prospect to become the
eleventh black-hole X-ray nova.  Thus we monitored the brightness of
the optical counterpart closely when it first appeared in the night
sky in late October and we found that it had returned to its
pre-outburst brightness (e.g. V = 19.0 on 2000 October 29.48 UT).  In
early December, using the new 6.5m MMT, we obtained the spectroscopic
observations detailed herein.

\section{OBSERVATIONS AND ANALYSIS}

Spectroscopic observations of XTE J1118+480 were obtained with the new
6.5~m MMT telescope at the F. L. Whipple Observatory on the nights of
2000 December 1 and 4 (UT).  The Blue Channel spectrograph was used
with the Loral CCD (3072~x~1024) detector and the 500 gpm grating.
This configuration yielded $\approx$~3.6~\AA\ resolution (FWHM) for a
slit width of 1.0$''$, which approximately matched the seeing on the
two observing nights.  The sky conditions were clear.  Two exposures
of XTE J1118+480 (900 s each) were obtained on December 1, and six
additional exposures (900--1200 s each) were obtained on
December~4. Immediately before and after each observation of the
object, an exposure was obtained of a wavelength calibration lamp
(He-Ne-Ar).  We also observed BD+12447, an M2 dwarf with a
well-determined systemic velocity, and the flux standard Feige~34.  In
our data analysis, we also made use of spectra of six additional dwarf
stars, which were obtained with precisely the same focal-plane
instrumentation in earlier MMT observing runs.  The wavelength
calibrations were interpolated dispersion solutions scaled according
to the time of an observation relative to the time of the lamp
exposures.  Cross correlations between the flux-calibrated spectra of
XTE J1118+480 and the template spectra of the G/M dwarfs were computed
for the range 4900--6500~\AA. The spectral reductions and the
cross-correlation analysis were performed using the software package
IRAF\footnote{ IRAF (Image Reduction and Analysis Facility) is
distributed by the National Optical Astronomy Observatories, which are
operated by the Association of Universities for Research in Astronomy,
Inc., under contract with the National Science Foundation.}.

Photometric monitoring observations were performed using the 1.8~m
Vatican Advanced Technology Telescope (VATT) located at the Mount
Graham International Observatory on the nights of 2000 November~30 and
December~1 (UT).  These observations were conducted using the VATT CCD
Camera and Loral CCD detector (2048 x 2048 pixels) and a Harris~I
filter ($\lambda_{0}$~=~8105~\AA; FWHM~=~1624~\AA).  The CCD was
binned at 2x2 pixels providing a scale of 0.4$''$/pixel.  Ninety-six
consecutive images were obtained on November 30 and an additional 53
images were obtained on December 1.  The typical integration time was
120~s and the time between consecutive observations was typically 3
minutes.  On both nights, seeing was 2$''$ at the start of the
observations due to high airmass, but quickly improved to 1$''$
thereafter.  There was light cirrus on the first night, and the second
night was clear.  Images were processed to eliminate the electronic
bias, correct for pixel-to-pixel sensitivity variations, and remove
significant interference fringes in the images.  The relative
intensities of XTE J1118+480 and selected field stars were computed by
using DAOPHOT photometry.  The photometric calibration was performed
using Landolt standard stars.

\centerline{
\psfig{figure=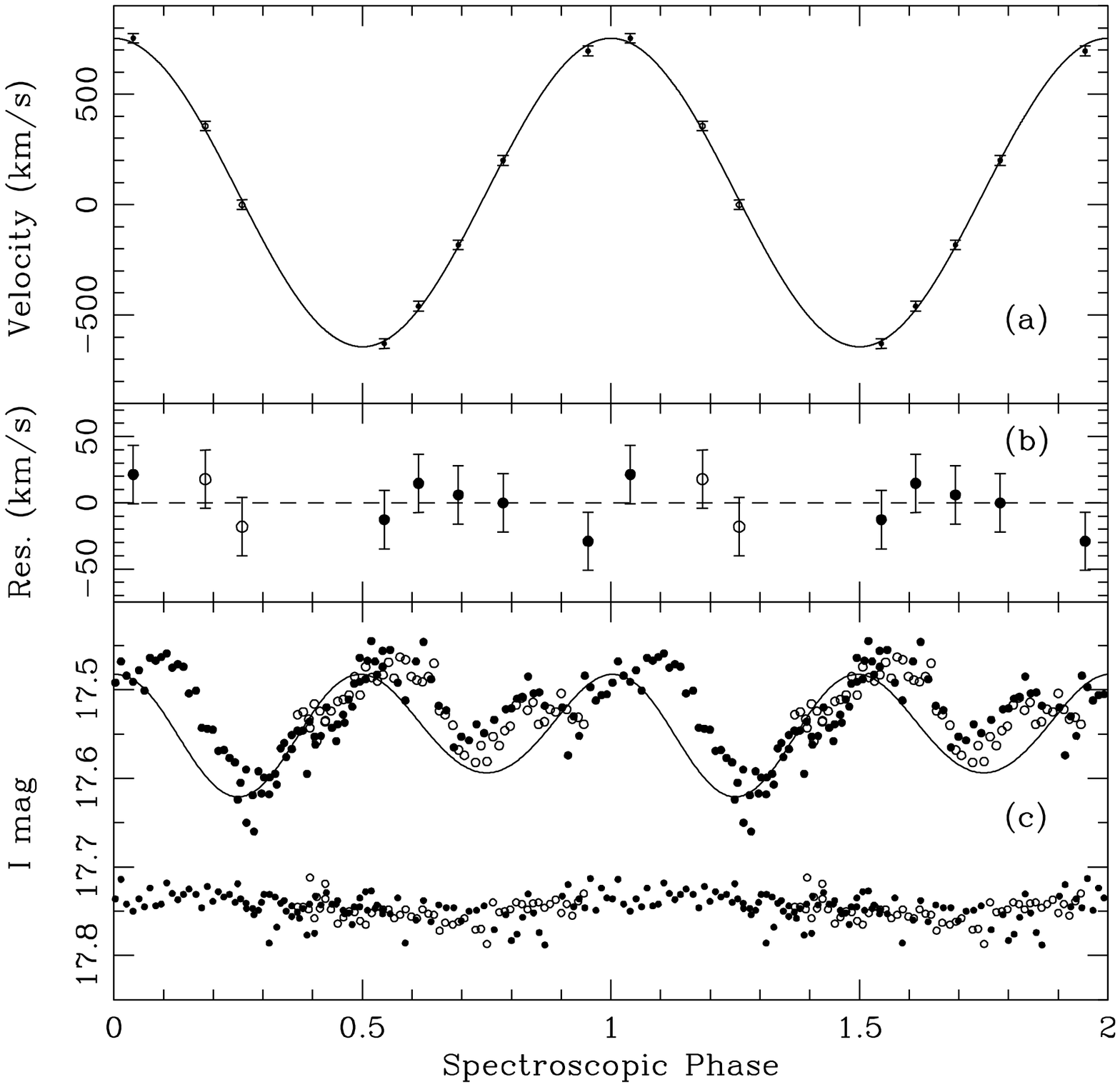,width=4.0in}
}
\centerline{
\begin{minipage}[h]{3.5in}
{\small Figure 1: {Spectroscopic and photometric data folded on the
orbital period and the ephemeris given in Table~1. (a) Radial velocity
measurements of the secondary star.  Smooth curve is a fit to a
circular orbit based on the velocity amplitude and phase given in
Table 1. (b) The residual differences between the data and the fitted
curve.  Open/filled symbols are for December~1/4, respectively.  (c)
Top trace is the I-band light curve of XTE J1118+480 with a superposed
ellipsoidal model.  Filled/open symbols are for November
30/December~1, respectively.  Lower trace shows the intensity of a
nearby comparison star of magnitude I = 17.0; the trace has been
offset by 0.75 mag for convenience of display.  The rms variation in
the star's intensity is 0.015 mag. \\}}
\end{minipage}
}

\section{SPECTROSCOPIC RESULTS}

We used the spectra of the velocity standards as cross-correlation
templates to derive a radial velocity curve for the secondary star.
The eight individual spectra of XTE J1118+480 were cross-correlated
against each of our seven template stars, which ranged in spectral
type from G8V to M2V.  Comparable velocity curves were obtained with
each template spectrum.  However, based on the Tonry \& Davis (1979) R
value, which is a measure of signal to noise achieved in a
cross-correlation, we found that the K3V, K5V and K5/8V templates
yielded the best correlations.  The eight velocities derived using the
K5/8V template star, GJ9698, are shown in Figure 1a.  The velocity
data imply that the secondary star in XTE J1118+480 is orbiting a
compact object with a velocity amplitude of approximately
700~km~s$^{-1}$.  The large velocities of the secondary contrast
sharply with the behavior of the night-sky lines, which show an rms
variation of less than 10~km~s$^{-1}$.  

Assuming a sine function, the velocities are well fit by the orbital
parameters given in Table 1, where $T_{\rm 0}$ is the time of maximum
velocity, $V_{\rm 0}$ is the systemic velocity, $K_{2}$ is the
velocity semiamplitude of the secondary, and $P$ is the orbital
period.  These four parameters were fit simultaneously using the
IDL routine {\it curvefit}.  A preliminary account of these
dynamical results (McClintock et al. 2001) and the consistent results
obtained by a second group (Wagner et al. 2001) appeared earlier in
the IAU Circulars.  In \S4 we argue that the period given in Table 1
is the correct orbital period, not an alias.  In fitting the
velocities, we have assumed that the eight velocity errors are all the
same because the R values are all comparably high ($\approx$~7-12).
We have adjusted this error to the value 24~km~s$^{-1}$ in order to
give $\chi_{\nu}^{2}$~=~1.0.  The orbital parameters in Table 1 define
the velocity ephemeris, which is represented by the solid line in
Figure~1a.  The post-fit residuals are shown in Figure 1b.  The mass
function may be derived from the above results:

$$f(M) \equiv {(M_{1}\ sin\ i)^{3} \over (M_{1} + M_{2})^{2}} =
{PK_{2}^{3} \over 2\pi G} = 6.00 \pm 0.36 M_{\odot}.$$

Since the mass of the compact primary necessarily exceeds the value of
the mass function, our results imply that the primary is much too
massive to be a neutron star within general relativity and is
therefore a black hole (Rhoades \& Ruffini 1974).

\small

\begin{center}
\begin{tabular}{p{1.5in}l}
\multicolumn{2}{c}{TABLE 1} \\
& \\
\multicolumn{2}{c}{SPECTROSCOPIC ORBITAL PARAMETERS}
\\
\hline \hline
\multicolumn{1}{c}{Parameter}&\multicolumn{1}{c}{Result} \\ \hline
$T_{\rm 0}$ (UT)$^a$\dotfill\ &2000 December 1.6476$\pm$ 0.0010 \\
$T_{\rm 0}$ (heliocentric)$^a$\dotfill\ &JD 2,451,880.1485$\pm$ 0.0010 \\
$V_{\rm 0}$ (km s$^{-1}$)\dotfill\ &26$\pm$17 \\
$K_{\rm 2}$ (km s$^{-1}$)\dotfill\ &698$\pm$ 14 \\
$P$ (days)\dotfill\ &0.17013$\pm$0.00010 \\
a$_{2}$\ sin\ $i$ (R$_{\odot})~^{b}$\dotfill\ &2.35$\pm$ 0.05 \\
$f(M/M_{\odot}$)\dotfill\ &6.00$\pm$ 0.36 \\ \hline
\multicolumn{2}{l}{$^{a}$Time of maximum redshift.} \\
\multicolumn{2}{l}{$^{b}$Projected orbital radius of the secondary.}
\end{tabular}
\end{center}

\normalsize

An average of the six spectra taken on December 4 in the rest frame of
the secondary star is shown in Figure 2a.  Before averaging the
individual spectra, they were Doppler shifted to zero velocity using
the velocities predicted by the ephemeris in Table 1.  The spectrum of
the template star, GJ9698, is shown in Figure 2b for comparison.  Most
of the stronger absorption lines of GJ9698 are evident in XTE
J1118+480.  The most prominent features are the continuum
discontinuity at Mg $b$ ($\sim$5175~\AA) and the Na~I~5890-96~\AA\
doublet. As noted above, the cross-correlation analysis favors
template stars of mid-K spectral type over those with spectral types
of M0V or later.  However, an inspection of the rest-frame spectrum
itself suggests that it is somewhat later than mid-K.  Given our
limited signal-to-noise, we conclude that the spectral type of the
secondary is in the range K5-M1.  Since the orbital period and mass
estimates imply a binary separation of $\sim$~3 R$_{\odot}$, the
secondary is presumed to be luminosity class V.

Because of the very low column depth to the source (\S1), the Na I
line is quite free of interstellar contamination.  We therefore use
its equivalent width to estimate the relative contributions of the
secondary star and the accretion disk to the total light at 5900~\AA.
For the spectrum of XTE J1118+480 (Figure 2a) we find EW =
$2.8~\pm~0.2$\AA.  For five comparison stars with spectral types
ranging from K5V to M1V, we find EW = 6.6--10.8~\AA.  From these
results we conclude that the K/M dwarf secondary contributes
34\%~$\pm$~8\% of the total light at 5900~\AA, a result we use in \S4
to analyze the ellipsoidal light curve and we now use to estimate the
distance to the source.

\centerline{
\psfig{figure=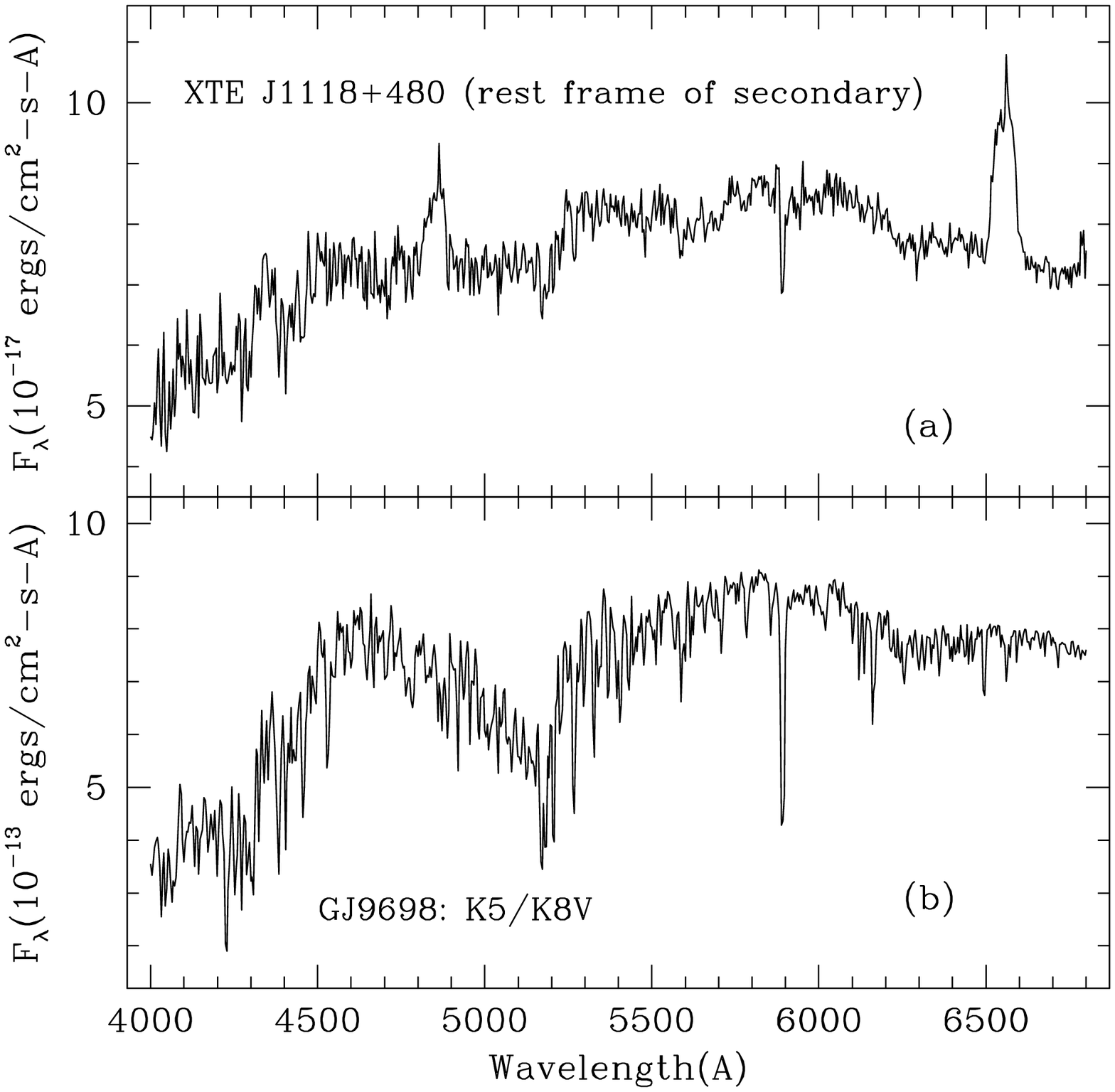,width=3.5in}
}
\centerline{
\begin{minipage}[h]{3.5in}
{\small Figure 2: {(a) Spectrum of XTE J1118+480 in the rest frame of
the secondary star.  The photospheric absorption features are most
apparent in this frame; however, the prominent Balmer emission lines
are significantly distorted.  (b) The spectrum of GJ9698, which was
used as a velocity template for the cross-correlation analysis.\\}}
\end{minipage}
}

We estimate the distance to XTE~J1118+480 using ``method~II''
described in Barret, McClintock, \& Grindlay (1996).  For the
secondary, we compute an average density of $\rho = 6.9$~g~cm$^{-3}$
from the orbital period and assume M$_{2}$~=~0.4~M$_{\odot}$, which is
very probably correct to within a factor of 2 (e.g. van Paradijs \&
McClintock 1994).  With these inputs we calculate
R$_{2}$~=~0.45~R$_{\odot}$.  We use the total magnitude of the optical
counterpart, V~=~19.0 (\S1), and the fraction of the light contributed
by the secondary, to estimate the magnitude of the secondary: V =
20.1~$\pm$~0.3.  Finally, for the range of spectral types in question,
K5V--M1V, we obtain an estimate of the distance:
d~=~1.8~$\pm$~0.6~kpc.  There are two nearly equal ($\sim$~25\%)
contributions to the error: the uncertainty in the spectral type of
the secondary, and the (assumed) factor of two uncertainty in the mass
of the secondary.

The spectrum of XTE J1118+480 shows strong Balmer lines, which
indicate the presence of an accretion disk.  In individual exposures,
the Balmer lines are often double-peaked and broad with widths in the
range 2300--2900~km~s$^{-1}$ (FWHM).

\section{PHOTOMETRY RESULTS}

The I-band light curve of XTE J1118+480 folded on the spectroscopic
ephemeris is shown in Figure 1c; data for a comparison star of
comparable magnitude are plotted just below.  The light curve shows
two maxima and two minima per orbital cycle.  This behavior is the
hallmark of an ellipsoidal light curve, which is commonly observed for
quiescent black-hole X-ray novae.  However, the light curve deviates
significantly from an ideal ellipsoidal light curve, which is
represented by the solid line (see \S5), in several ways.  For
example, there is extra light near phase 0.8, which can be explained
as due to the bright spot (Warner 1995).  A more problematic deviation
from the ellipsoidal model is the apparent phase lag of the light
curve relative to the spectroscopic ephemeris. Fitting the light curve
to a sinusoid gives a phase lag of 0.050 $\pm$ 0.008, which
corresponds to a time delay of 12.2 $\pm$ 2.0 minutes.  In contrast,
studies of other quiescent X-ray novae indicate good agreement between
the photometric and spectroscopic phases (e.g. McClintock \& Remillard
1986; Shahbaz et al. 1994; Orosz \& Bailyn 1997).  Consequently, this
12-minute phase lag calls into question the ellipsoidal nature of the
light curve.  Possibly XTE J1118+480 was not yet fully quiescent
during our observations, even though our dynamical results (Table 1)
are entirely consistent with those obtained more than a month earlier
by Wagner et al. (2001).  Possibly our light curve is dominated by
eclipse effects, which can effectively shift the phase of a light
curve.  An example of this phenomenon is the set of light curves
observed for GRO J1655-40 as it approached quiescence in early 1995
(see Fig. 1 in Bailyn et al. 1995).  Future observations in deep
quiescence can be expected to resolve these issues.  In the meantime,
we assume below in \S5 that the light curve is ellipsoidal.

Could this 12-minute phase difference be due to an error in the data
clock at either the VATT or the MMT?  We believe that the answer to
this question is ``no'', despite the fact that the performance of
these clocks was not rigorously checked at the time of the
observations.  The time base for both observatories is Network Time
Protocol (NTP) via SUN computers, which routinely provides reliable
and precise time to these observatories.  Moreover, both observatories
also use their NTP connection for the precise ($\sim$~1'') pointing of
their telescopes.  If the NTP-based time had been in error by even 10
s during the observations, the telescope operator and observer would
have noted gross errors in the telescope pointing; none was observed.
Finally, independent and simultaneous photometry of XTE J1118+480 was
obtained by P. Groot using the FLWO 1.2m telescope; the light curve
derived from these data agrees in phase with our VATT light curve to
within 0.010 in phase or 2.4 minute in time.  We conclude that a
terrestrial origin of the 12-minute phase offset appears very
unlikely, and we believe that the offset is due to the source itself.

We searched the photometric data shown in Figure 1c for periodicities
by computing the variance statistic of Stellingwerf (1978) for trial
periods between 0.01 d and 0.5 d.  Deep minima in the
$\Theta$-statistic occur only at $P_{phot}$ = 0.170 $\pm$ 0.006~d and
at half that period.  The statistical uncertainty in the period
determination was estimated using a Monte Carlo method (Silber et
al. 1992).  The adopted {\it spectroscopic} period is approximately
$P~=~0.1701~\pm~0.0001$~d (Table~1) and its two closest aliases are
$P_{-}$ = 0.1610 d and $P_{+}$ = 0.1803 d.  We now give four reasons
for rejecting these alias periods and adopting $P$ = 0.1701~d as the
orbital period: (1) The light curves obtained by folding the
photometric data on $P_{-}$ and $P_{+}$ are complex and much less
compelling than the light curve shown in Figure 1c, as expected since
they differ from the best photometric period by $>~1.5\sigma$. (2) A
superhump modulation (Warner 1995) was repeatedly observed during
outburst; its period decreased from 0.1708 $\pm$ 0.0001~d (Patterson
2000) to 0.1703 $\pm$ 0.0001~d (Uemura et al. 2000) over the course of
several weeks.  These results argue very strongly in favor of $P$ =
0.1701 $\pm$ 0.0001~d and against the aliases (e.g. Bailyn 1992; Kato,
Mineshige \& Hirata 1995). (3) The $T_{0}$ given by Wagner et
al. (2001) agrees with our $T_{0}$ for $P$ = 0.1701 $\pm$ 0.0001~d,
but disagrees if one adopts $P_{-}$ or $P_{+}$.  (4) Wagner et
al. (2001) independently found $P$ = 0.1699~$\pm$ 0.0001~d with
spectroscopic observations separated by 10 nights.  We therefore
conclude that $P$ = 0.17013~$\pm$ 0.00010~d is the correct orbital
period.

\section{ON THE MASS OF THE BLACK HOLE}

We now use the absence of X-ray eclipses and a preliminary analysis of
the light curve to further constrain the mass of the black hole.
Despite very extensive X-ray observations of XTE J1118+480 in
outburst, no X-ray eclipses have been reported.  We can use this
result to place an upper limit on the inclination angle, which boosts
somewhat the 6.0 M$_{\odot}$ mass limit that is set by the mass
function.  We consider two models for the secondary: (1) An
0.5~M$_{\odot}$ star with a radius of 0.5~R$_{\odot}$ that just fills
its Roche lobe.  In this case we find that an absence of eclipses
implies i~$<$~$79.5^{\rm~o}$ and M$_{1}$~$>$~7.2~M$_{\odot}$.  (2) A
very low-mass secondary, M$_{2}$~=~0.2~M$_{\odot}$, which we assume
just fills its Roche lobe radius of 0.35~R$_{\odot}$.  In this case,
we find i~$<$~$81.8^{\rm o}$ and M$_{1}$~$>$~6.5~M$_{\odot}$.  Here we
have used the mean radius of the Roche lobe in calculating the eclipse
condition.

We modeled the I-band light curve (Fig. 1c), which we assume to be
ellipsoidal (but see \S4), using a computer code written by Yoram Avni
(1978; see also Orosz \& Bailyn 1997).  We assumed a K7V stellar
atmosphere, a limb darkening coefficient of u = 0.60 (Al-Naimy 1978),
and a gravity darkening exponent of $\beta$ = 0.08.  We assumed that
the star fills its Roche lobe and that its rotation period is the
orbital period.  We further assumed that M$_{1}$/M$_{2}$~=20, although
the light curve is very insensitive to the choice of this parameter
for M$_{1}$/M$_{2}~\gtrsim~10$.  The biggest uncertainty is the fraction
of the light at 8100~\AA\ that is non-stellar; we call this component
the ``disk fraction.''  For the purposes of this approximate analysis,
we assume that the disk fraction at 8100~\AA\ is 66\%, the same as the
value we derived at 5900~\AA\ in \S3.  We computed a set of
ellipsoidal models for the star for i~=~$40^{\rm o}$ to i~=~$90^{\rm
o}$ in steps of $5^{\rm o}$.  To each model, we added a constant
component of the flux corresponding to the 66\% contribution of the
accretion disk.

The model that best matches the folded light curve is shown in Figure
1c.  This model corresponds to a very high inclination,
i~=~$80^{\rm~o}$, and a value for the mass of
M$_{1}$~=~7.2~M$_{\odot}$ (for M$_{2}$~=~0.5~M$_{\odot}$).  This
result is consistent, but just barely, with the limits obtained above
from the absence of X-ray eclipses.  There are several caveats on this
preliminary analysis, the most important of which concerns our use of
the disk fraction at 5900~\AA\ as a proxy for the unknown disk
fraction at 8100~\AA.  The available evidence indicates that the disk
fraction decreases with increasing wavelength (e.g. Oke 1977; Casares
et al. 1993; Marsh, Robinson, \& Wood 1994).  Consequently, we have
very likely added too much disk light to our models.  As a
hypothetical example, consider the effect of adding a disk fraction of
only 40\% (instead of 66\%) to our models: In this case we would have
found i~=~$52^{\rm o}$ and M$_{1}$~=~13.2~M$_{\odot}$ (for
M$_{2}$~=~0.5~M$_{\odot}$).  A very strong upper limit on the mass is
obtained by making the extreme assumption that the disk contributes no
light at all in the I band.  In this case we find i~$>~40^{\rm o}$ and
M$_{1}$~$<$~24~M$_{\odot}$.  Despite the overriding uncertainty in the
I-band disk fraction, our provisional light curve results suggest that
the orbital inclination is relatively high, i~$\gtrsim~55^{\rm o}$,
and that the black hole mass is correspondingly modest
M$_{1}~\lesssim$~10~M$_{\odot}$.  The broad Balmer emission lines
(\S3) also suggest a high inclination.

\section{CONCLUSION}

With XTE J1118+480 there are now eleven X-ray novae that have been
dynamically confirmed to contain black hole primaries.  For
XTE~J1118+480 we find an exceptionally large mass function,
$6.00~\pm~0.36$ M$_{\odot}$, which is rivaled only by that of V404 Cyg
(Casares, Charles, \& Naylor 1992).  XTE~J1118+480 is additionally
distinguished by having the shortest orbital period (4.08 hr) of the
black hole binaries.  Finally, the extraordinarily low column depth
(N$_{H} \approx 1.3 \times 10^{20}$~cm$^{-2}$) and modest distance
($\sim$~1.8 kpc) of XTE~J1118+480 make this system central to the
study of Galactic black holes.

\smallskip

We are grateful to Paul Groot for help in confirming the photometric
phase and to Mike Fitzpatrick and Frank Valdes for IRAF support. This
work was supported in part by NASA through grant DD0-1003X and
contract NAS8-39073.

\end{document}